\documentclass[11pt]{article}
\usepackage[dvips]{graphicx}
\usepackage[utf8]{inputenc}
\usepackage[english]{babel}
\usepackage[T1]{fontenc}

\usepackage[title,titletoc]{appendix}
\usepackage[nottoc]{tocbibind}

\usepackage{color}
\usepackage{float}
\usepackage{tabularx}
\usepackage{indentfirst}
\usepackage{hyperref}
\usepackage{pdfpages}
\usepackage{fancybox}
\usepackage{setspace}
\usepackage{amsmath}
\usepackage{listings}
\usepackage{wrapfig}
\usepackage{chngcntr}

\usepackage{makecell}

\newenvironment{capalayout}{
    \setlength{\topmargin}{-3.0cm}
    \setlength{\footskip}{0cm}
    \thispagestyle{empty}
} 

\newcommand{\ISSNno}{0103-9741}
\newcommand{\MCCSeqAno}{05/2021}
\newcommand{\TituloCapa}{Building a Noisy Audio Dataset to Evaluate Machine Learning Approaches for Automatic Speech Recognition Systems}

\newcommand{\AutorANome}{Julio Cesar Duarte}
\newcommand{\AutorAemail}{jduarte@inf.puc-rio.br}
\newcommand{\AutorBNome}{Sérgio Colcher}
\newcommand{\AutorBemail}{colcher@inf.puc-rio.br}

\usepackage{amsthm}

\theoremstyle{definition}

\usepackage{listings}
\usepackage{caption}
\usepackage{subcaption}
\usepackage[table,xcdraw]{xcolor}

\begin{document}
\begin{capalayout}
        \centering
        \includegraphics[keepaspectratio,width=14.7cm]{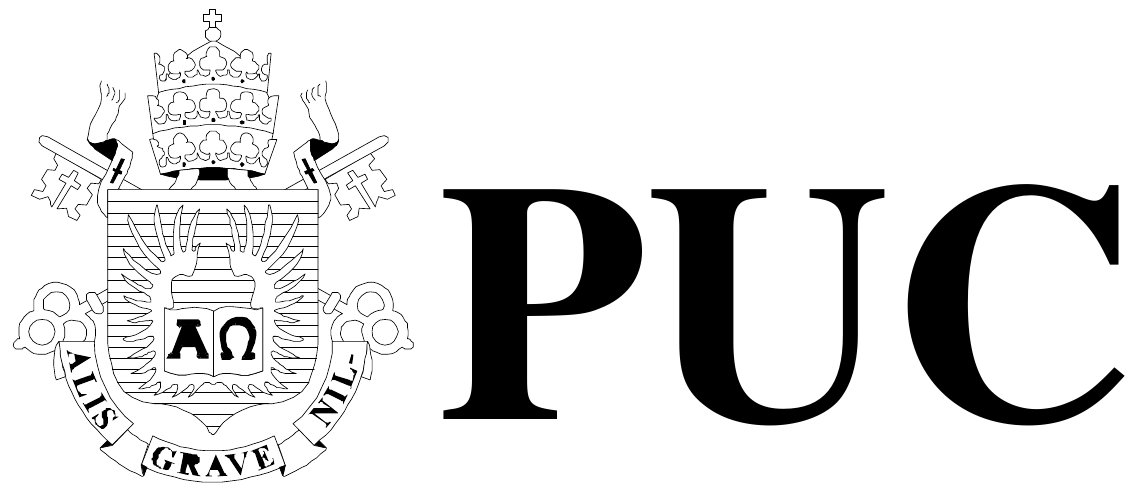}
        
        \medskip
    
        \setlength\fboxsep{1pt}
        \shadowbox{\fbox{\begin{minipage}[h]{14.5cm}
            \begin{center}
                \doublespacing
                \fontfamily{phv} \fontsize{14}{16} \selectfont
                \medskip
                ISSN \ISSNno
                
                \bigskip
                Monografias em Ciência da Computação 
                
                n\textordmasculine \, \MCCSeqAno
                
                \bigskip
                \medskip
                \fontsize{18}{20}\selectfont
                \textbf{\TituloCapa}
                
                \bigskip
                \fontsize{14}{15}\selectfont
                \textbf{\AutorANome} \\
                \textbf{\AutorBNome} 

                \bigskip
                
                \medskip
                Departamento de Informática
                \bigskip
            \end{center}
        \end{minipage}}}
        
        \bigskip
        \bigskip
        
        \begin{minipage}[h]{14.5cm}
            \doublespacing
            \fontfamily{phv}
            \begin{center} 
                \fontsize{12}{14}\selectfont
                \textbf{PONTIFÍCIA UNIVERSIDADE CATÓLICA DO RIO DE JANEIRO} \\
                \textbf{RUA MARQUÊS DE SÃO VICENTE, 225 - CEP 22451-900} \\
                \textbf{RIO DE JANEIRO - BRASIL}
            \end{center}
        \end{minipage}
        \newpage
    \end{capalayout}
\thispagestyle{empty}

\begin{flushleft}
\begin{tabular}{p{11.1cm}r}
Monografias em Ciência da Computação, No. \MCCSeqAno & ISSN: \ISSNno \\
Editor: Prof. Carlos José Pereira de Lucena   & Setembro, 2021
\end{tabular}
\end{flushleft}
\LARGE
\begin{center}
    {\bf \TituloCapa}
\end{center}
\normalsize
\begin{center}
{\bf \AutorANome~and \AutorBNome}
\end{center}
\begin{center}
   \AutorAemail, \AutorBemail
\end{center}

\noindent {\bf Abstract.} Automatic speech recognition systems are part of people's daily lives, embedded in personal assistants and mobile phones, helping as a facilitator for human-machine interaction while allowing access to information in a practically intuitive way.
Such systems are usually implemented using machine learning techniques, especially with deep neural networks. 
Even with its high performance in the task of transcribing text from speech, few works address the issue of its recognition in noisy environments and, usually, the datasets used do not contain noisy audio examples, while only mitigating this issue using data augmentation techniques.
This work aims to present the process of building a dataset of noisy audios, in a specific case of degenerated audios due to interference, commonly present in radio transmissions. 
Additionally, we present initial results of a classifier that uses such data for evaluation, indicating the benefits of using this dataset in the recognizer's training process.
Such recognizer achieves an average result of 0.4116 in terms of character error rate in the noisy set (SNR = 30).
\medskip

\bigskip

\noindent {\bf Resumo.} Sistemas de reconhecimento automático de fala fazem parte da vida cotidiana das pessoas, embutidos em assistentes pessoais e telefones celulares, auxiliando como um facilitador para a interação homem-máquina ao mesmo tempo que permitem o acesso à informação de forma praticamente intuitiva.
Tais sistemas são normalmente implementados utilizando técnicas de aprendizado de máquina, especialmente com redes neurais profundas.
Mesmo com o seu alto desempenho na tarefa de transcrever o texto da fala, poucos trabalhos levam em consideração a questão do seu reconhecimento em ambientes ruidosos e, normalmente, as bases utilizadas não apresentam exemplos de áudios com ruídos, apenas mitigando tal questão, por intermédio de técnicas de aumentação de dados. 
Este trabalho tem por objetivo apresentar o processo de construção de uma base de dados de áudios ruidosos, em um caso específico de áudios degenerados em função de interferências, comumente presentes em transmissões rádio. 
Adicionalmente, apresentamos resultados iniciais de um classificador que utiliza tais dados para avaliação, indicando os benefícios de se utilizar este conjunto já no processo de treinamento de um reconhecedor.
Tal reconhecedor alcança um resultado médio de 0.4116 em termos de taxa de erro de caracteres no conjunto ruidoso (SNR = 30).
\medskip

\noindent {\bf Palavras-chave:} Aprendizado de Máquina, Conjuntos de Dados, Sistemas de Re\-co\-nhe\-ci\-men\-to Automático de voz, Áudios Ruídosos.
bigskip

\newpage
\pagenumbering{roman} \setcounter{page}{2}
\vspace*{\fill}
\begin{flushleft}
    \textbf{In charge of publications:} \\
    PUC-Rio Departamento de Informática - Publicações \\
    Rua Marquês de São Vicente, 225 - Gávea \\
    22451-900 Rio de Janeiro RJ Brasil \\
    Tel. +55 21 3527-1516 Fax: +55 21 3527-1530 \\
    E-mail: publicar@inf.puc-rio.br \\
    Web site: http://bib-di.inf.puc-rio.br/techreports/ \\
    \end{flushleft}

\newpage
\renewcommand{\contentsname}{Table of Contents}
\renewcommand{\appendixname}{Annex}
\tableofcontents

\newpage
\pagenumbering{arabic} \setcounter{page}{1}
\counterwithin{lstlisting}{section}

\section{Introduction}
\label{sec:intro}

Automatic Speech Recognition (ASR) systems are composed of tools and techniques that allow machines to understand the human voice through speech, allowing them to mimic a common human interaction called conversation.
It is a field of computer science that has been researched for decades and consists mainly of deriving a speech transcript of an audio waveform from an usually digitized signal \cite{HELANDER1988301}.

\begin{figure}[!ht]
    \centering
    \includegraphics[width=.7\textwidth]{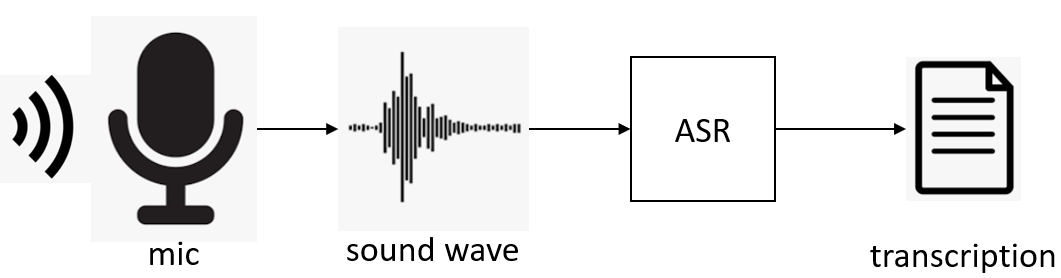}
    \caption{ASR application pipeline}
    \label{fig:asr}
\end{figure}

Figure~\ref{fig:asr} shows the outline of the application of an ASR.
Usually a microphone, or an array of them, is used to capture speech, digitizing its contents in a wave file, that can be fed to the software that will process it.
Finally, this software's output consists of the speech's transcription, that may be stored in a text file.

An ASR module, in any system, is very convenient for data entry, access to remote information, and interactive services such as \textit{chatbots}.
ASR modules also help handicapped people, allowing or helping them interact with systems in a way that could be hard or nearly impossible for them \cite{Juang2000AutomaticRA}.

Even with such practical and noble objectives, and having achieved high performance in their task in normal controlled environments, reliably transcription systems in realistic environments are still a challenge and their research and product development have been progressing for more than 30 years \cite{HELANDER1988301}.

As an important part within the concept of Natural Language Processing \cite{Abney:1991}, this task is now embedded in several equipment used daily, such as mobile phones and personal assistants and even with the excellent performance on the task obtained by the machine learning derived classifiers usually being used, a problem persists, speech recognition in noisy environments.
Normally, such classifiers use datasets with noise-free audio that are built with good quality microphones in controlled ambiences.
This reality may not be found in real-world noisy environments, such as, for example, when engines are present or in radio transmissions.

Current ASR systems often rely on online services such as Google Cloud, Microsoft Azure, and IBM Watson Speech To Text APIs \cite{xu2021benchmarking}.
They typically perform well in controlled environments and can be constantly updated. 
However, as they are online services, they depend on a stable Internet connection, in addition to having an operating cost.

There are also other offline solutions, such as CMU Sphinx \cite{lamere03thecmu}, which are limitedly updated and with less support for languages other than English. 
Among the \textit{generic} solutions that allow the development of an ASR for any language, DeepSpeech \cite{HannunCCCDEPSSCN14} stands out, by allowing, through the training of deep neural networks implemented using the Tensorflow framework, the development of an ASR for any language by just feeding the tool with a dataset in the desired language, in addition to adjusting its hyper-parameters. 
Deepspeech has even proven performance close to the state of the art for both English and Mandarin \cite{amodei16}.

Although ASRs prove to be a very important task in NLP, most of the developments are based on the English language, although some recent works are investigating the performance of such techniques for other languages, such as Portuguese \cite{gris2021bresci, gris2021brazilian, quintanilha2020asr}.

Likewise, ASRs can be used in a military environment, incorporated in a solution for audio transcription on a military radio, which usually suffers from a lot of interference.
In this context, there is the Software-Defined Radio Project \cite{rds2020}, which today is Brazilian Army's priority 1.1 \cite{brasil2019}, where radio families with such technology could automatically identify the audio that is received (or retransmitted), presenting the corresponding text on its display, in addition to storing the information for future utilization.
A speech recognition technology that enables its use in noisy environments, such as a high frequency (HF) communication channel, commonly used in the Brazilian theater of operations, would be, then, extremely useful.

The main objective of this work is to present the methodological process of building a noisy audio dataset that allows the evaluation of ASRs derived from machine learning algorithms. 
As a case study, radio transmission paths on HF channels will be used, especially, but not limited, for use in military operations, with the Portuguese language.
As an additional contribution, we will also present initial experiments with the generated dataset that illustrate the importance of adapting the ASRs to the real environment in which they are used.

In the remainder of this article, we will present additional steps toward the construction of this dataset.
First, in Section~\ref{sec:related}, we show related work, focusing on Portuguese audio datasets, as well as works that deal with ASRs in noisy environments. 
Section~\ref{sec:dataset} details the construction process showing the assumptions we took, as well as the simulation tools used to represent the chosen environment. 
We, then, conclude this article in Section~\ref{sec:experiments} with our experimental results that validates the dataset that is followed by our final remarks and possible future work in Section~\ref{sec:conclus}.
\section{Related Work}
\label{sec:related}

ASRs are a hot topic for research in computer science and many papers can be reviewed in order to help build custom systems, but only a small amount deals with noise in audio.
Conversely, few works are related to the Portuguese language, mainly because, but not limited to, the limited availability of voice speech data with their validated transcriptions \cite{gris2021bresci}.
If we consider both restrictions, the Portuguese idiom and noisy data, this limits even more the comparison with similar articles. 
Nevertheless, this section summarizes the most related works found in our research.

\subsection{Portuguese datasets for speech recognition applications}
\label{sec:relds}

The commonly called dataset CETUC \cite{alcaim2008lsf} was the first initiative found for a Portuguese-speaking dataset derived from the CETEN-folha text corpus.
It consists of 145 hours of noise-free speech text from 100 speakers repeating once the same 1,000 sentences, or 3,528 words.
These phrases were chosen from CETEN-folha, having phonetic balance as their main characteristic \cite{ciri2005conjunto}.
All recordings were made in a studio where the audio was captured at a sampling rate of 16 kHz and 16 bits per sample.

LibriSpeech \cite{pratap2020largescale} is a multilingual dataset benchmark for speech related tasks.
It comprises of about 50.5K hours of speech spread in eight languages.
The Portuguese subset of Librispeech contains approximately 170 hours of voice from 62 different speakers. 
The dataset was derived from several audiobooks downloaded from the Internet.

The Common Voice Dataset \cite{ardila-etal-2020-common} is a multilingual open-source project to build voice datasets for basically any language.
Anyone, through the project's site\footnote{https://commonvoice.mozilla.org/}, can record or listen to and evaluate a sentence, in order to contribute to the project.
Its current version 7.0 supports more than 75 different languages.
The Portuguese subset comprises 3GB of over 84 hours of validated speech and approximately 2,000 different speakers.
This work uses Common Voice as its base dataset because of two major factors.
First, this is an ongoing project that everyone can contribute to, and as a result, the project is expected to have good longevity.
Second, this is the project officially supported by the DeepSpeech tool (Section~\ref{sec:experiments}), also used for our ASR experiments.

Table~\ref{tab:datasets} summarizes the main characteristics of the datasets depicted here. 
While these datasets can be used to build well-performing ASRs, they are nowhere near the size when compared to similar datasets in English.
Furthermore, they are built with audio files that were recorded in a controlled environment and do not represent possible noisy environments in which the systems can be employed.

\begin{table}[ht]
\centering
\caption{Portuguese voice datasets main characteristics}
\begin{tabular}{|l|c|c|l|}
\hline
\rowcolor[HTML]{EFEFEF} 
\textbf{Dataset} & \textbf{Total Hours} & \textbf{Speakers} & \textbf{Audio Source} \\ \hline
CETUC \cite{alcaim2008lsf} & 145 & 100 & studio recording \\ \hline
LibriSpeech \cite{pratap2020largescale} & 170 & 62 & audiobooks \\ \hline
Common Voice \cite{ardila-etal-2020-common} & 84 & 2,000 & on-line recording \\ \hline
\end{tabular}
\label{tab:datasets}
\end{table}

\subsection{Speech recognition with noisy audio}
\label{sec:relnoise}

Oliveira Santos\cite{debora01ruido} presents a comparison between three techniques for speech recognition in noisy environments applied to a limited set of Portuguese words corresponding to the digits of the alphabet. 
Three types of noise were applied: white noise, chitchat and background industry noise.
Of the techniques that were used, the one that applies the Maximum Likelihood Linear Regression (MLLR) obtained the best results in conjunction with Hidden Markov Models (HMM).

Menêses Santos\cite{rafael16abordagem} analyzes the use of a hybrid model of Convolutional Neural Networks (CNN) and HMM in conjunction with noisy audios for speaker-dependent speech recognition.
The experiments were performed on audio datasets that contain utterances and digits.
The results presented with noises that represent chitchat, engine and industry noises, using a signal-to-noise ratio (SNR) of 6db, showed that the use of this hybrid model improves the quality of the system even without the explicit incorporation of a noise suppression technique.

Prodeus et al.\cite{prodeus16training, prodeus17automatic} show that the use of noisy audio samples during the training phase of an ASR can improve its performances. 
In their work, several types of noise were used, such as those coming from grinders, computers, trucks, among others, in a range of SNR from 0 to 40 db. 
Different techniques were used where the signal-to-noise ratios and the types of noise were combined in the training and test (recognition) sets.

Wang et al.\cite{wang18twostage} proposes the use of Deep Neural Networks (DNNs) as a way to improve the performance of an ASR that was already trained using only noiseless audios. 
The system was evaluated using a dataset of 21,800 English utterances with added noise ranging from a SNR of 5 to 20 db.
By applying the technique in conjunction with an ASR trained using a 6-layer context-dependent (CD)-DNN-HMM hybrid model with 2048 nodes per layer, a word error rate of 6.56\% was obtained from a system that already performed at 17.89\%.

Ribeiro\cite{ribeiro19reconhecimento} investigates the use of neural networks in the development of ASRs using a throat microphone, which is a more suitable device for capturing audio in noisy environments.
The system uses audio samples of simple commands (``OK'' and ``Cancel'') and digits for the Portuguese Language, which were noise filtered prior to being applied to a classification system composed by an ensemble of three classifiers: Multilayer Perceptron, Binary Multilayer Perceptron and Self-Organizing Maps. 
The performance of 96.6\% in the hit rate indicates that the use of the device can help with voice recognition in noisy environments.

Maruf et al.\cite{maruf20effects} analyzes the effects of noise in ASRs using two feature extraction techniques called Mel Frequency Cepstral Coefficients (MFCC) and Relative Spectral Transform-Perceptual Linear Prediction (RASTA-PLP). 
The proposed CNN-derived ASR was trained using digits and 11 utterances of the Bangla language from 120 different speakers. 
Urban noises were added from different environments, like libraries, street and rail stations.
The best results of 93.18\% in terms of accuracy were found using the RASTA-PLP model.

Finally, Pervaiz et al.\cite{pervaiz20incorporating} investigates the use of data augmentation to improve noise robustness of ASRs.
It uses several public utterance datasets to assess the quality of different machine learning approaches.
The approaches tested were traditional deep neural networks such as Deep Feed-Forward Networks, Long Term Memory Networks (LSTMs) and CNNs that were evaluated on a noise dataset based on an Asian accent.

In summary, the effects of noise on ASRs are a common research topic.
However, few works deal with all aspects of ASRs, such as any type of transcription, and none propose a methodology for building datasets to be used for the purpose of evaluating machine learning techniques.
Furthermore, only a few focus on the Portuguese language.
In this sense, this work intends to fill this gap by proposing a process of building noisy audio datasets to be used in training ASRs. 
In addition, a specific use case for the Portuguese language is implemented, where we present some results of proposals inspired by the related works reported here.
\section{Building the Dataset} 
\label{sec:dataset}

In this section, we present the steps taken in our proposal to build a noisy audio dataset.
This process is generic and can be customized depending on the type of noise you want to model in the ASR.
First, it is necessary to start with the relationship between noise and the original signal that is used, since it plays an important role in the dataset construction process.

In our methodology, we can vary the signal-to-noise ratio (SNR) when combining or simulating the chosen environment with a noiseless audio dataset. 
SNR is a metric for comparing signal power to noise and provides an important indication of the degree to which the
the signal was contaminated with additive noise \cite{carlson02comm}.
if we want to mathematically formalize, SNR is the quotient (ratio) between the mean power of a signal and the power of the background noise. 
Equation \ref{eq:snr} illustrates the mathematical formula to evaluate the SNR of any signal.

\begin{equation}
    SNR = 20 \cdot log \frac{RMS(s)}{RMS(n)}
    \label{eq:snr}
\end{equation}

where RMS is a function that evaluates the square root of the averages of data being analyzed, the signal (s) or noise (n).

When noise has an approximately Gaussian distribution, it is called an Additive White Gaussian Noise (AWGN) and it is the most commonly assumed model \cite{carlson02comm}.

In our dataset generation approaches, we use the following SNRs as input: \{-30, -20, -10, -5, 0, 5, 10, 20, 30\}. 
The positive sub-interval of this range was chosen because it is a set of values widely used in similar works. 
However, we made sure to also include the same negative sub-interval, in which the noise is stronger than the signal itself.
Such range can be interesting in works that perform audio pre-processing for noise extraction, although they are not very useful in machine learning approaches, since, when listening to an audio from these subsets, even a person is not able to perform its transcription.

In order to build a speech corpus with noisy audio, we first need a noise-free corpus.
In our process, we use the Common Voice 6.1 dataset
\cite{ardila-etal-2020-common} (Section~\ref{sec:relds}).
The main reason for this choice is the fact that it is a dataset with active and collaborative development, keeping the release of new versions every 6 months, approximately.
Furthermore, this set is officially supported by the DeepSpeech platform, used in our experiments.
More details of DeepSpeech will be explained in Section~\ref{sec:experiments}.

Our case study of interest is voice transmission over a military analog radio transmission channel. 
In Brazil, one of the most used transmission bands is HF, due to its wide use in areas such as the Amazon.
Thus, scenarios are created that allow the simulation of this channel in an audio (voice) transmission.
We created a total of four scenarios to run this simulation, each with its own degree of reliability.
These four scenarios, which can be replicated for different case studies, are explained in the next subsections.

\subsection{1st Scenario - White Noise}
\label{sec:scenario1}

Our first scenario is a simple AWGN model where we have a randomly generated signal with equal strength at different frequencies with a flat spectral density.
This is the simplest model, since it does not emulate a particular noise, but can be used as a generic approach for modeling noise in any channel.

Adding this kind of noise to a signal is simple and can be done with the Python snippet presented in Listing~\ref{lst:whitenoise}.

\clearpage

\lstset{language=Python, caption = An excerpt of a python code that adds white noise to a signal}
\label{lst:whitenoise}
\begin{lstlisting}
def addWhiteNoise(signal, SNR):
    RMSs = sqrt(mean(signal**2))    
    RMSn = sqrt(RMSs**2/(pow(10, SNR/10)))
    noise = random.normal(0, RMSn, signal.shape[0])
    return signal + noise
\end{lstlisting}

\begin{figure}[!ht]
    \centering
    \includegraphics[width=0.4\textwidth]{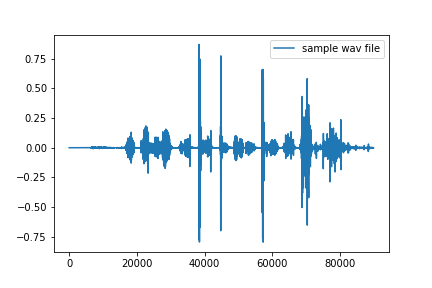}
    \caption{Sample .wav file without noise}
    \label{fig:wavnonoise}
\end{figure}

\begin{figure}[ht]
     \centering
     \begin{subfigure}[b]{0.3\textwidth}
         \centering
         \includegraphics[width=\textwidth]{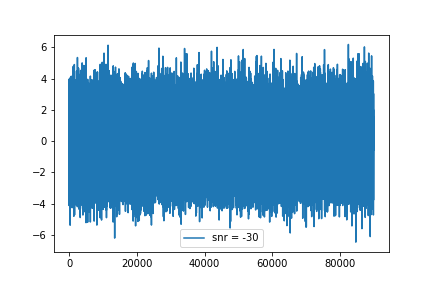}
         \caption{$SNR=-30$}
         \label{fig:SNR-30}
     \end{subfigure}
     \hfill
     \begin{subfigure}[b]{0.3\textwidth}
         \centering
         \includegraphics[width=\textwidth]{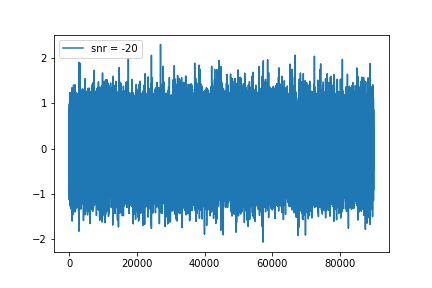}
         \caption{$SNR=-20$}
         \label{fig:SNR-20}
     \end{subfigure}
     \hfill
     \begin{subfigure}[b]{0.3\textwidth}
         \centering
         \includegraphics[width=\textwidth]{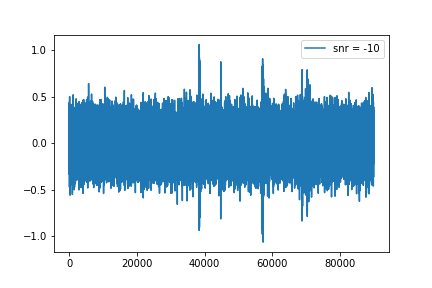}
         \caption{$SNR=-10$}
         \label{fig:SNR-10}
     \end{subfigure}
\newline        
     \begin{subfigure}[b]{0.3\textwidth}
         \centering
         \includegraphics[width=\textwidth]{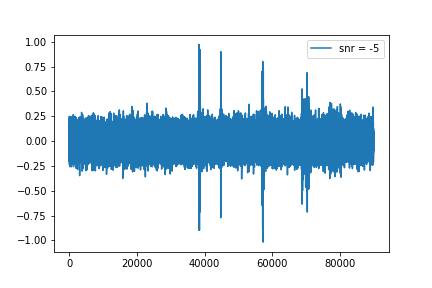}
         \caption{$SNR=-5$}
         \label{fig:SNR-5}
     \end{subfigure}
     \hfill
     \begin{subfigure}[b]{0.3\textwidth}
         \centering
         \includegraphics[width=\textwidth]{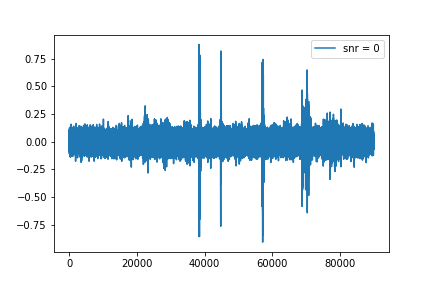}
         \caption{$SNR=0$}
         \label{fig:SNR0}
     \end{subfigure}
     \hfill
     \begin{subfigure}[b]{0.3\textwidth}
         \centering
         \includegraphics[width=\textwidth]{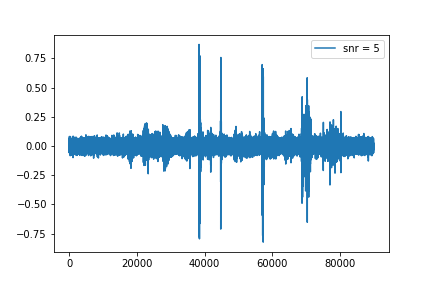}
         \caption{$SNR=5$}
         \label{fig:SNR5}
     \end{subfigure}
\newline
     \begin{subfigure}[b]{0.3\textwidth}
         \centering
         \includegraphics[width=\textwidth]{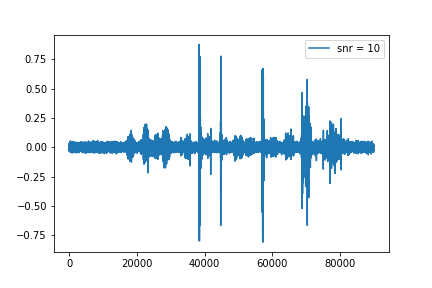}
         \caption{$SNR=10$}
         \label{fig:SNR10}
     \end{subfigure}
     \hfill
     \begin{subfigure}[b]{0.3\textwidth}
         \centering
         \includegraphics[width=\textwidth]{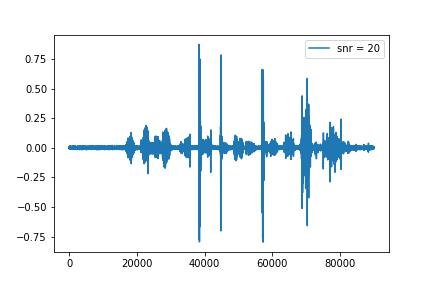}
         \caption{$SNR=20$}
         \label{fig:SNR20}
     \end{subfigure}
     \hfill
     \begin{subfigure}[b]{0.3\textwidth}
         \centering
         \includegraphics[width=\textwidth]{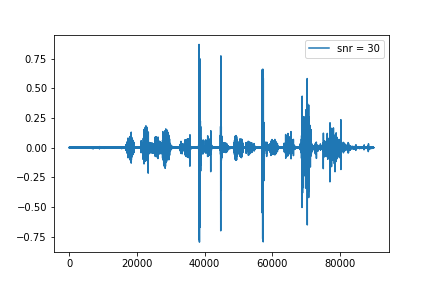}
         \caption{$SNR=30$}
         \label{fig:SNR30}
     \end{subfigure}

        \caption{AWGN applied to a sample .wav file with different SNR values}
        \label{fig:awgnsnr}

\end{figure}

Note that the SNR parameter influences noise degradation, since the lower the value (higher in terms of negative value), the more degraded the signal will be. 
This can be better observed in Figure~\ref{fig:awgnsnr} that depicts the application of different SNR values in the sample wav file shown in Figure~\ref{fig:wavnonoise}.

Observe that the higher the SNR value (Figure~\ref{fig:SNR30}), the more similar are the obtained signal and the original one.

\begin{figure}[!ht]
    \centering
    \includegraphics[width=0.4\textwidth]{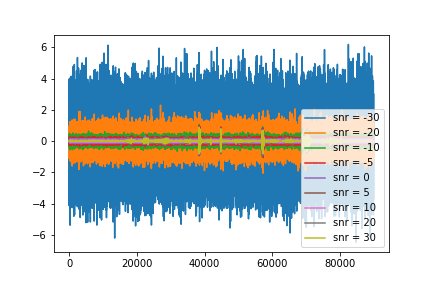}
    \caption{AWGN applied to a sample .wav file with all SNR values}
    \label{fig:wavawgn}
\end{figure}

Finally, we show in Figure~\ref{fig:wavawgn} all generated files, in order to compare them within the same scale.

\subsection{2nd Scenario - Noise Collected from Live Antennas}
\label{sec:scenario2}

Although AWGN is the most used scenario to model generic noise, it does not replace the use of noise found in a communication channel.
Thus, models trained with white noise may not perform well when applied in the real world.
Using a more reliable representation of the noise in the environment, in this case, leads to ASRs with a better performance.

An easy way to generate audio with more representative noises is to collect this noise directly on the communication channel that the system is used. 
Such captured noise can then be merged with the original audio files, generating a dataset closer to the reality of applying the recognizer.

\begin{figure}[!ht]
     \centering
     \begin{subfigure}[b]{0.32\textwidth}
         \centering
         \includegraphics[width=\textwidth]{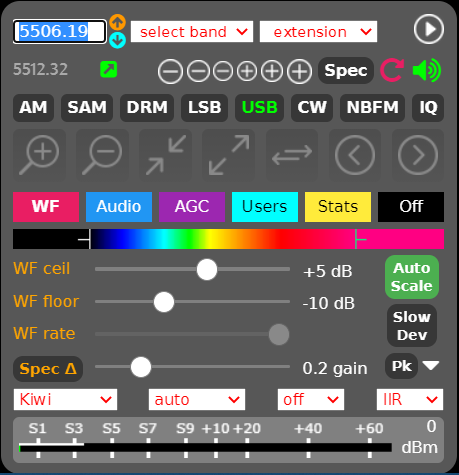}
         \caption{Radio Web Interface}
         \label{fig:interface}
     \end{subfigure}
     \hfill
     \begin{subfigure}[b]{0.45\textwidth}
         \centering
         \includegraphics[width=\textwidth]{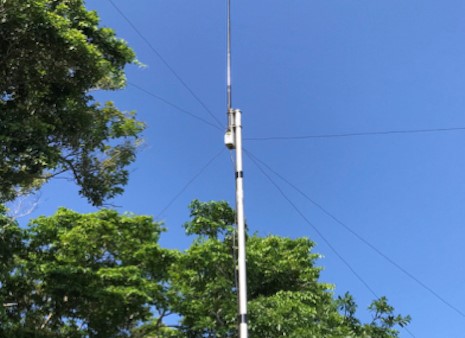}
         \caption{Antenna's site. Source: \cite{py1eme}}
         \label{fig:antenna}
     \end{subfigure}
        \caption[Equipment used to sample noise files]{Equipment used to sample noise files}
        \label{fig:py1eme}
\end{figure}

In order to accomplish this idea, we used a remote HD radio service provided by Halász \cite{py1eme}.
Through this site, we control a multi-user SDR receiver software with a web interface, called OpenWebRX\footnote{https://github.com/ha7ilm/openwebrx} that uses two Kiwi SDRs connected to dipoles (40m and 80m) and vertical (20m) antennas.
The site of these antennas is located in Ilha dos Ratos, Paraty, Rio de Janeiro. Figure~\ref{fig:py1eme} illustrates both the radio control interface (\ref{fig:interface}) and the antenna's site (\ref{fig:antenna}).

While using this setup, we sampled receptions on the following frequencies~/~bands: 500~KHz~/~AM, 1 MHz~/~AM, 1,836 MHz~/~USB, 8,060 MH~/~CW and 7,794 MHz~/~USB. 
These frequencies were chosen since they were active during the collection period, although they were recorded when no transmission was detected.

Those recording samples form a pool of noise sources which can then be applied to any audio file by, randomly, choosing an excerpt of this noise pool.
A Python snippet of these method is presented in Listing~\ref{lst:filenoise}.

\lstset{language=Python, caption = An excerpt of a python code that adds a random cut of noise stored in a file to a signal}
\label{lst:filenoise}
\begin{lstlisting}
def addNoise(signal, noiseSource, SNR):
    signal = interp(signal, (signal.min(), signal.max()),
        (-1, 1)) 
    startCut = randrange(0, len(noiseSource) - len(signal))
    noise = noiseSource[startCut:startCut + len(signal)]
    RMSs = sqrt(np.mean(signal**2))        
    RMSn = sqrt(RMSs**2/(pow(10, SNR/10)))
    noise = noise*(RMSn/sqrt(mean(noise**2)))
    signal_n = signal + noise
\end{lstlisting}

\begin{figure}[!ht]
    \centering
    \includegraphics[width=0.4\textwidth]{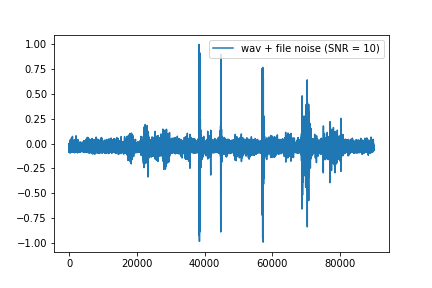}
    \caption{Sample .wav with added noise from samples (SNR = 10)}
    \label{fig:wavfilenoise10}
\end{figure}

For instance, we show the application of this method in Figure~\ref{fig:wavfilenoise10}, with a SNR value of 10. 
The same sample .wav file depicted in Figure~\ref{fig:wavnonoise} was also used as input.

\subsection{3rd Scenario - Software Simulation Noise}
\label{sec:scenario3}

Using collected files is a good strategy to represent a communication channel when you do not have the implementation of a mathematical model for the communication channel, however, such files can limit the possibilities of noise used in the training of a speech recognizer.
Thus, when using a software that implements such models, the variability of the nuances of the noise being analyzed is increased.

\begin{figure}[!ht]
     \centering
     \includegraphics[width=.5\textwidth]{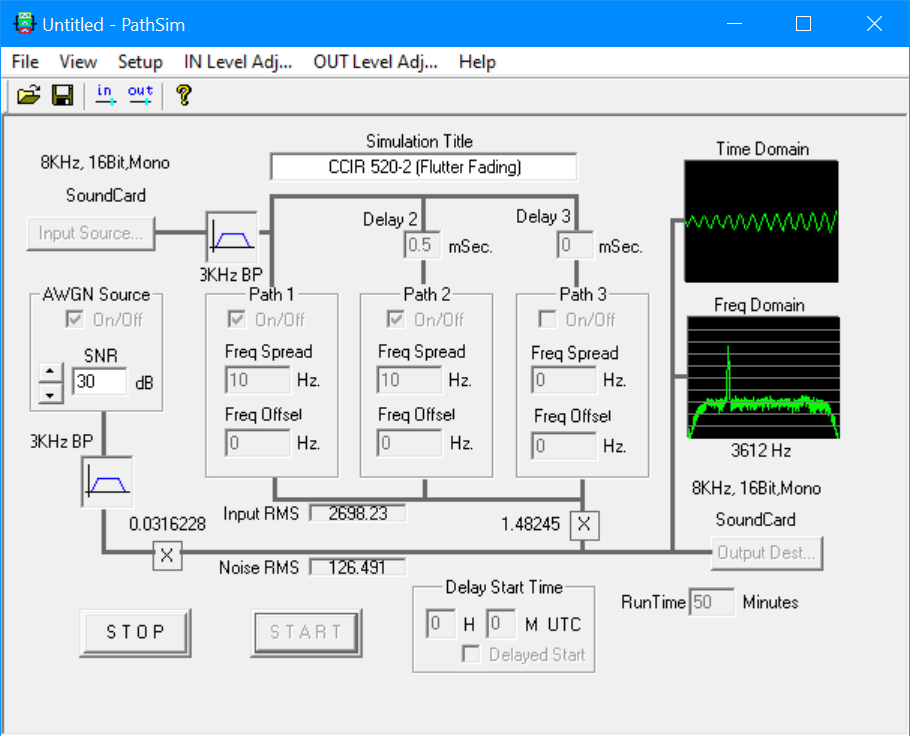}
     \caption{PathSim Interface}
     \label{fig:pathsim}
\end{figure}

PathSim is a software that simulates HF radio propagation paths using an audio file or the computer's sound card \cite{pathsim}.
Three HF paths are available with different spreading frequencies parameters and frequency offset functions. 
The spreading uses a complex AWGN signal to achieve the desired SNR which is filtered by a 3KHz low pass filter to limit the noise. 
Figure~\ref{fig:pathsim} shows PathSim's interface, while the sound card input was injected with a 1 KHz sine wave test tone.

Initially, the intention was to process all audio files directly in PathSim, however some limitations prevented this approach.
First, the software does not provide a programmable API, so it was necessary to manually feed the files into its interface.
To get around the large amount of files needed to be processed, they were merged into files of maximum 500 minutes, PathSim's limit.
However, sometimes the software simply crashed when processing these big .wav files.
It is important to note that this is older software from the 2000s, without active development, running on a new operating system.
As a result, smaller files were generated and manually fed into the software.
With these files, initial tests were performed on the recognizer, which detected problems in the generated .wav files.
Some of these files were corrupt and some of which produced incompatible results.
Finally, the software has the limitation of working only with 8 KHz .wav files, which generates files with worse quality and that need to be resampled after processing.

In order to circumvent these problems, a different approach was adopted, similar to that of Scenario 2 (Section~\ref{sec:scenario2}).
Several noise files were generated using the software and, then, were merged to the noise-free dataset using the same code in Listing~\ref{lst:filenoise}.

In our study case, we use the \textit{CCIR-Flutter} preset, which presented the most degenerative noises, following the International Radio Consultative Committee - CCIR 520-2 \cite{ccir5202}.
In this configuration, two paths are used with the same attenuation, differential time delay of 0.5 ms and a Doppler spread of 10.0 Hz with zero Doppler shift. 
Other setups can also be used.

\begin{figure}[!ht]
    \centering
    \includegraphics[width=0.4\textwidth]{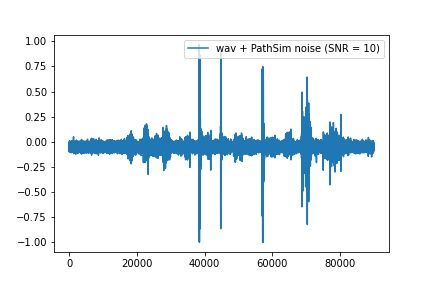}
    \caption{Sample .wav with added noise from PathSim (SNR = 10)}
    \label{fig:wavpathsim10}
\end{figure}

With this approach, we maintain the advantage of using a custom mathematical model for the communication channel, even with its software limitations.
Figure~\ref{fig:wavpathsim10} shows the application of this approach, with a SNR value of 10, when merged to the sample .wav file represented in Figure~\ref{fig:wavnonoise}.

\subsection{4th Scenario - Hardware Simulation Noise}
\label{sec:scenario4}


Although the strategy used in Section~\ref{sec:scenario3} allowed the generation of audio files with customized noise through a mathematical model of the communication channel, we also choose to use a more robust solution, embedded in a dedicated hardware.
A dedicated hardware unit is usually built for a single purpose and, therefore, excels at the task for which it was manufactured.

\begin{figure}[!ht]
     \centering
     \begin{subfigure}[b]{0.32\textwidth}
         \centering
         \includegraphics[width=\textwidth]{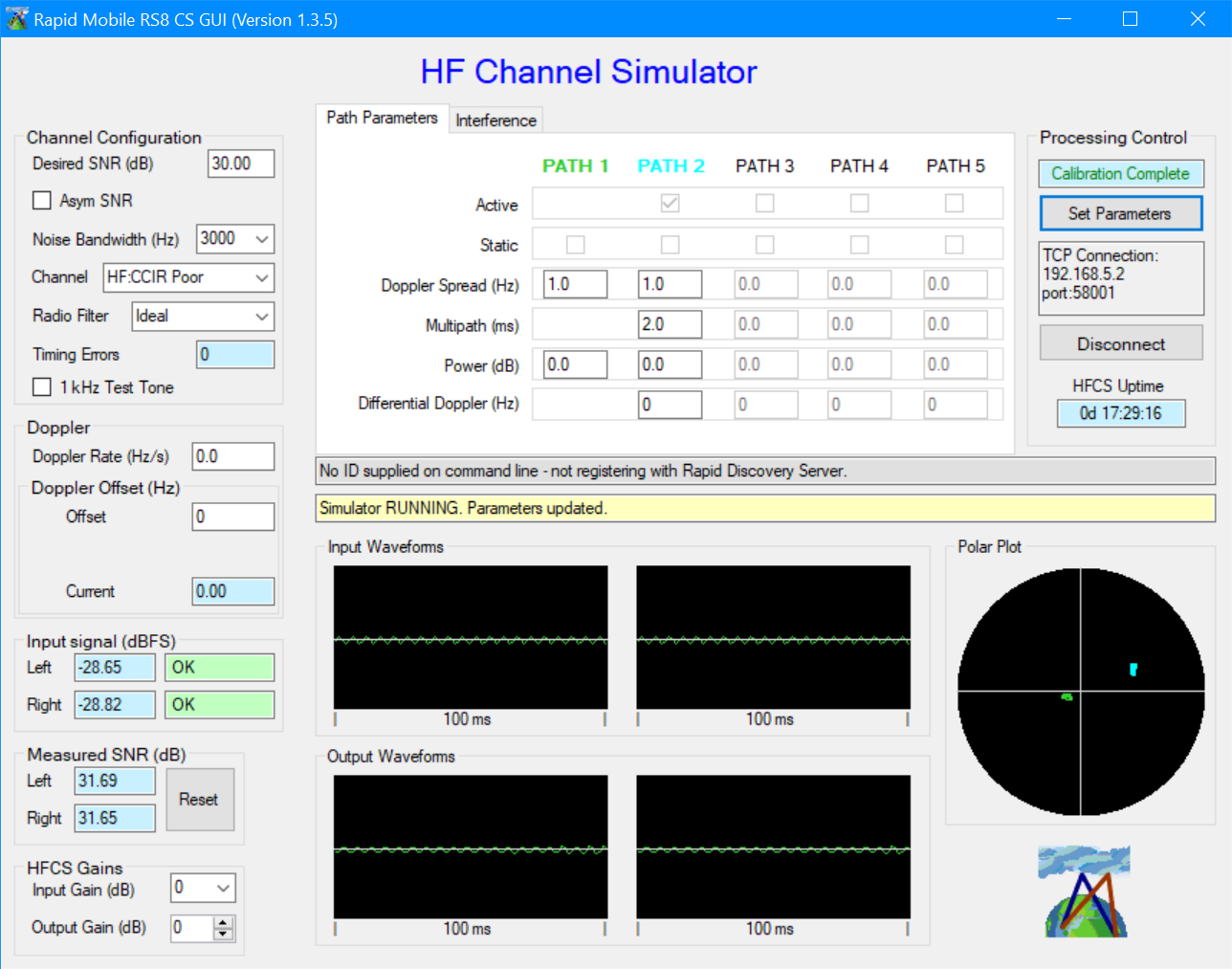}
         \caption{Application Interface}
         \label{fig:interfacers8}
     \end{subfigure}
     \hfill
     \begin{subfigure}[b]{0.45\textwidth}
         \centering
         \includegraphics[width=\textwidth]{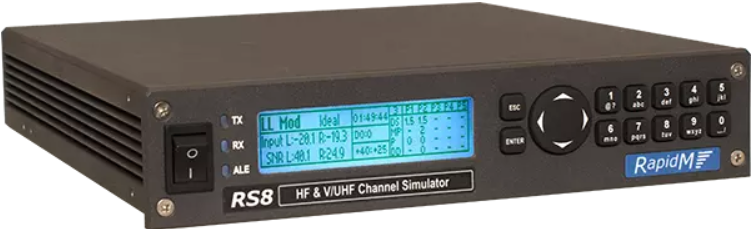}
         \caption{RS8 device. Source: \cite{rapidmrs8}}
         \label{fig:equipmentrs8}
     \end{subfigure}
        \caption[RS8 Hardware and Software]{RS8 Hardware and Software Interfaces}
        \label{fig:rapidmrs8}

\end{figure}

With that idea in mind, RapidM RS8 was used.
The RS8 is a device that allows the simulation of HF and V/UHF channels in order to evaluate modems and waveforms at baseband which is in compliance with military standards like MIL-STD-188-110B  \cite{rapidmrs8}.
Figure~\ref{fig:rapidmrs8} presents the software and hardware interface of the equipment.

Using a device such as RS8 allows us to achieve the state-of-the-art in terms of simulation of communication channels, thus representing it more reliably.
However, a disadvantage here is present. 
Unlike the software solutions presented in Sections \ref{sec:scenario1}, \ref{sec:scenario2} and \ref{sec:scenario3}, all processing is performed on a external equipment, requiring the audio to be sent to the computer's sound card output, which is connected to the device through a customized cable, processed by the equipment, and, finally, captured in one of the inputs of the same sound card.
All this takes place in real time, that is, for every second of audio in the dataset, one second of processing is required and the process is repeated for all the signal-to-noise ratios used.

Similar to Section~\ref{sec:scenario3}, and to give some noise variance, we configured the device with the \textit{CCIR-Poor} preset, also recommended in CCIR 520-2 \cite{ccir5202}, which has two independently fading paths, but with a differential time delay of 2 ms and a Doppler frequency spread of 1.0. 
As with PathSim, other setups may be used.

\begin{figure}[!ht]
    \centering
    \includegraphics[width=0.4\textwidth]{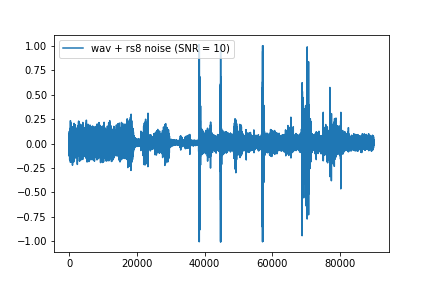}
    \caption{Sample .wav with added noise from RS8 (SNR = 10)}
    \label{fig:wavrapidm10}
\end{figure}

Using this equipment, we obtain .wav files like the one depicted in Figure~\ref{fig:wavrapidm10}, using a SNR value of 10.
It is noteworthy that no file merge is done in this case, as all processing is done in real time.

\subsection{Final considerations in building the dataset}

In this section, we present four ways to generate a noisy audio dataset.
This 4-step methodology is generic and can be applied to any kind of noise.
In our case study, we specialize it for noises generated in HF communication channels.
These semi-synthetic datasets can be used in the training phases of machine learning algorithms, either alone, if they well reflect the noise being modelled, or together, feeding the algorithms with more examples, as in a specialized data augmentation scheme.

All script and noise files used are available in the project's GitHub repository\footnote{ \url{https://github.com/duartejulio/noisy-asr}} allowing the replication of the experiments, with the exception of the RS8 output files, which can be provided on request.

\section{Dataset Validation}
\label{sec:experiments}

In order to evaluate the dataset in terms of its application potential for training with machine learning algorithms, we conducted some experiments on the construction of an ASR for Portuguese using the four approaches presented in Section~\ref{sec:dataset}.

In our experiments, we use
Mozilla DeepSpeech \cite{HannunCCCDEPSSCN14, deepspeech}, which is an open source speech-to-text framework that can use models trained by machine learning algorithms implemented within Google's TensorFlow \cite{tensorflow2015whitepaper}.
DeepSpeech has a flexible architecture where you can try different hyperparameters to build ASRs for any specific language.
It also supports the Common Voice Dataset, which makes our noisy datasets easier to use.

\begin{figure}[!ht]
    \centering
    \includegraphics[width=0.7\textwidth]{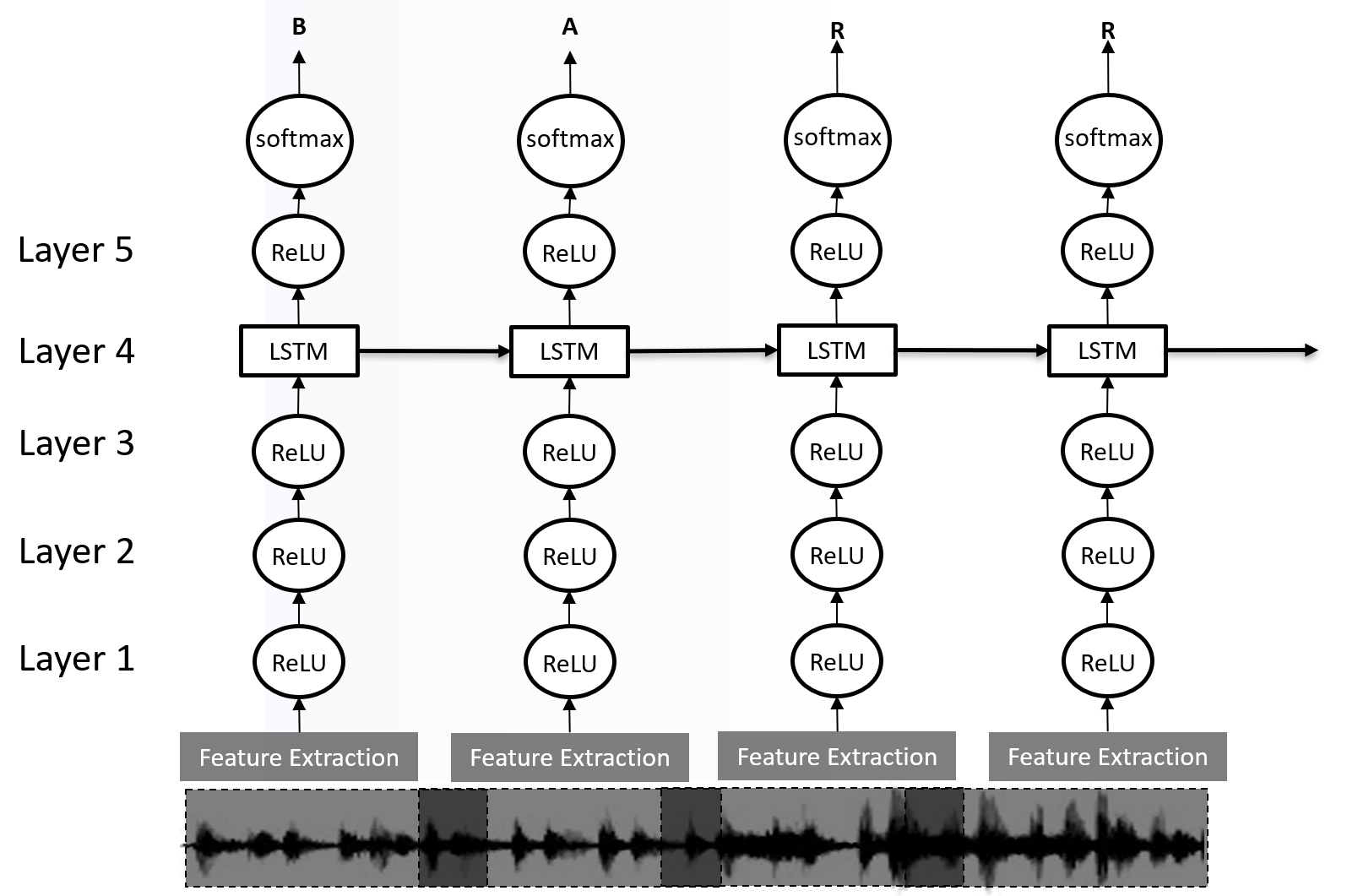}
    \caption[DeepSpeech basic architecture]{DeepSpeech basic architecture. Adapted from \cite{deepspeech}}
    \label{fig:rnnarchitecture}
\end{figure}

Deespech uses a Recurrent Neural Network architecture like the one depicted in Figure~\ref{fig:rnnarchitecture}. 
It consists of 5 hidden layers, where the first three and the fifth are composed of regular neurons and use a clipped rectified-linear (ReLu) activation function.
The fourth layer is a recurring LSTM layer that includes a set of hidden units with forward recurrence.
Finally, the output layer uses the softmax function which outputs the probabilities of the characters of the alphabet.

For our experiments, we use the following hyperparameters in order to train the neural networks: $epochs = 100$, $batch\_size=16$, $number\_of\_neurons\_per\_layer = 2048$, $feature\_extraction\_audio\_window\_length=32 ms$, $learning\_rate = 0.0001$, and $dropout\_rate = 0.40$. 
All other hyperparameters were set to the default values.


With this hyperparameter setup, we conducted two experiments. 
The first one uses only the train and dev subsets of Common Voice combined for training the neural networks. 
The dev subset is usually used for validation purposes, but in order to have a bigger training set, we opted to use it in conjunction to the train subset. 
Although Common Voice has about 50 hours of validated voice, its subsets are built in a way to ensure that the same speakers do not appear in different sets, nor that the same sentences appear repeated in the subsets. 
This considerably reduces the availability of data for training so, by combining the train and dev subsets, we get a total of 11,086 training examples. 

\begin{figure}[!ht]
    \centering
    \includegraphics[width=0.7\textwidth]{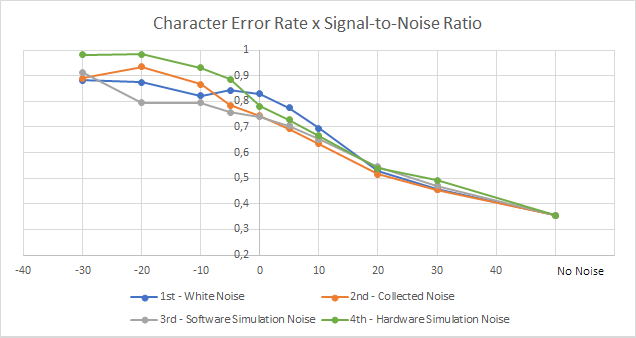}
    \caption{CER results for the model trained with the train+dev sets}
    \label{fig:results1}
\end{figure}

Figure~\ref{fig:results1} shows the results of this model in the same test set but applied to all four approaches presented in Section~\ref{sec:dataset} and all considered SNR values.
In this graph, we use character error rate (CER) as our main metric for evaluation.

First, we can see that, for the negative SNR values, it is very hard for the models to learn anything.
The ability to generalize of the models is very low, with erratic results. 
Normally, such results indicate a massive classification only of the most frequent letter of the alphabet (\textquotesingle a\textquotesingle).
Next, it is possible to see the difference in terms of CER among the different approaches when comparing with the same SNR.
Finally, it seems that the approach with hardware simulation noise is the hardest to recognize.

We show here, only the results in terms of CER, but the ones considering WER have similar behavior.
We do not consider this comparison because evaluations with WER usually contemplate the use of a language model, in the output of the ASR, after the classification process. 
A language model assigns probabilities to translation hypotheses in the target language \cite{talbot07randomised}, and,
in this context, it is used to correct transcriptions both at character and word level. 
As we would like to observe the behavior of different forms of noise in the final transcripts of the text, we chose to disable the external scorer, in the output of DeepSpeech, that comprises the language model being used, as such model could mask these results.

In our second experiment, we used an expanded version of the train+dev combined subset.
As previously reported, Common Voice has over 50 hours of validated text, but only a small portion is used on those sets. 
We propose to use all versions of the same sentences spoken by different speakers in the same subset. 
As a result, we add all repeated sentences from the validated set, as long as they were spoken by a speaker from the same subset. 
In this way, it is possible to expand the combined set without such subsets having examples with sentences or speakers from the test subset. 
The final expanded train+dev combined set has a total of 28,068 training examples, more than doubling the amount of examples available for training. 

\begin{figure}[!ht]
    \centering
    \includegraphics[width=0.7\textwidth]{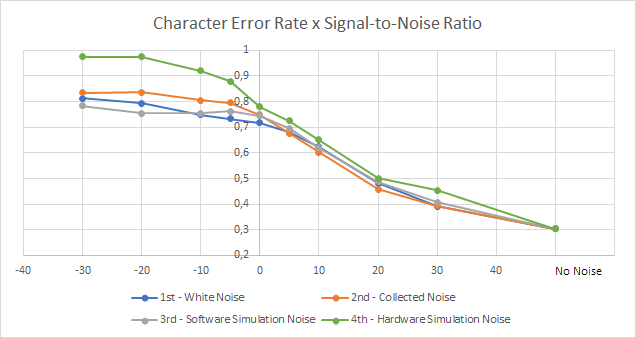}
    \caption{CER results for the model trained with the expanded train+dev sets}
    \label{fig:results2}
\end{figure}

Figure~\ref{fig:results2} shows the results with this new model.
It is important to note that the test subset has not changed from previous results.

Analyzing the results of Figure~\ref{fig:results2}, while comparing with the same results of Figure~\ref{fig:results1}, we observe that the new model has a better overall performance.
Note, nonetheless, that the results for the 4th approach, the one that uses hardware simulation noise, are still the worst, indicating that using a noisy audio dataset with those characteristics can improve such results.


Finally, it is not the intention of this work to generate an automatic speech recognizer for Portuguese, but rather to promote the discussion of the use of noisy audio data in its construction. 
However, it is interesting to show the possibilities of applying noisy audio datasets in a state-of-the-art speech recognizer.

Quintanillha et al.\cite{quintanilha2020asr} presents an ASR derived from DeepSpeech with a performance of 10.49\% and 25.45\%, in terms of CER and WER, respectively.
Also, Gris et al.\cite{gris2021brazilian} presents an ASR with a WER of 12.79\%.
The models used, however, are much more advanced, trained on larger datasets and use other strategies, such as language models, differently from this work.
Nonetheless, such works can be improved using the noise generation strategy proposed, especially if applied in environments with frequent noise.
Furthermore, as the proposal presented here is language-independent, it is possible to apply it in other languages under the same conditions.
\section{Conclusions and Future Work}
\label{sec:conclus}

Automatic speech recognition systems are currently part of our lives embedded in several hardware and software.
Such systems are generally derived from machine learning algorithms that require a large amount of data representative of the environment in which they are applied, in this case, speech samples with their respective transcriptions.
Often, such systems can be used in environments with constant and characteristic noise, such as noise from urban areas, industries, or even radio communication channels.

The main objective of this work is to propose a methodology for the generation of noisy audio datasets to be used in conjunction with machine learning algorithms in order to make them more robust for application in noisy environments.
In this sense, four approaches are proposed, each with distinct characteristics and representing differently the noise being added to the data.
Depending on the type of noise to be modeled and the availability of the needed tools, one or more approaches can be used together in order to provide more examples for the algorithm responsible for training the ASR, working as a specialized data augmentation strategy.
The evaluation of such sets in straightforward ASRs implemented points out that their use in the training phase can improve their performance, especially when applying the ASR in an noisy environment.

The main contribution of this work, in addition to the proposed methodology, is the generated dataset, as well as all the codes available for reproducing the construction of the dataset and applying it in a new context.
The codes can be downloaded from \url{https://github.com/duartejulio/noisy-asr}, as well as the noise samples collected during this work.

Even with the fulfillment of the work objectives, future works are always foreseen in order to improve or expand the results presented here.
With this purpose in mind, We also intend to use the noisy dataset in the ASR training phase, thus allowing a better evaluation of its behavior in the presence or not of the characterized noise.
In addition, we also intend to carry out tests with other types of noise comparing the performance of the proposed methodology with traditional and generic data augmentation schemes.
Finally, studies of how ASRs can be embedded in more compact hardware are also planned as future works, since such systems usually require a lot of computational power and the devices in which they can be installed do not always support such processing and memory requirements.

\section*{Acknowledgments}
This material is based upon work supported by Air Force Office Scientific Research under award number FA9550-19-1-0020.

We would also like to thank \textit{Centro Tecnológico do Exército} (CTEx) and \textit{Indústria de Material Bélico do Brasil} (IMBEL) for providing the means for us to use the RS8 equipment.

\bibliographystyle{plain}

\bibliography{bib}

\end{document}